\documentclass[10pt,aps,prd,floats,floatfix,preprintnumbers,superscriptaddress,nofootinbib]{revtex4-2}

\usepackage{graphicx}
\usepackage{epstopdf} 
\usepackage{dcolumn}
\usepackage{amssymb}
\usepackage[normalem]{ulem}
\usepackage{amsmath}
\usepackage{color}
\usepackage{xcolor,colortbl}
\usepackage[utf8]{inputenc}
\usepackage{slashed}
\usepackage[hidelinks]{hyperref}

\newcommand{\be}{\begin{equation}}
\newcommand{\ee}{\end{equation}}
\newcommand{\bes}{\begin{equation}\begin{split}}
\newcommand{\ees}{\end{split}\end{equation}}
\newcommand{\rme}{\mathrm{e}}
\newcommand{\rmd}{\mathrm{d}}
\newcommand{\rmi}{\mathrm{i}}

\newcommand{\calM}{\mathcal{M}}

\newcommand{\calP}{\mathcal{P}}
\newcommand{\calO}{\mathcal{O}}
\newcommand{\calD}{\mathcal{D}}

\newcommand{\xp}{\boldsymbol{x}_{\perp}}
\newcommand{\Phpv}{\boldsymbol{P}_{h\perp}}
\newcommand{\Php}{P_{h\perp}}

\newcommand{\yp}{\boldsymbol{y}_{\perp}}
\newcommand{\zp}{\boldsymbol{z}_{\perp}}
\newcommand{\qp}{\boldsymbol{q}_{\perp}}
\newcommand{\qonp}{\boldsymbol{q}_{1\perp}}

\newcommand{\kp}{\boldsymbol{k}_{\perp}}

\newcommand{\rp}{\boldsymbol{r}_{\perp}}
\newcommand{\Sp}{\boldsymbol{S}_{\perp}}
\newcommand{\aap}{\boldsymbol{a}_{\perp}}
\newcommand{\bp}{\boldsymbol{b}_{\perp}}

\newcommand{\Dp}{\boldsymbol{\Delta}_{\perp}}

\newcommand{\qop}{\boldsymbol{q}_{1\perp}}

\newcommand{\pd}{\partial}
\newcommand{\tr}{\mathrm{Tr}}

\date{January 27, 2025}

\begin{document}
\preprint{ZTF-EP-25-01}
\title{Single spin asymmetry in forward \ensuremath{pA} collisions from Pomeron-Odderon interference}

\author{Sanjin Beni\' c}
\author{Eric Andreas Vivoda}
\affiliation{Department of Physics, Faculty of Science, University of Zagreb, Bijeni\v cka c. 32, 10000 Zagreb, Croatia}

\begin{abstract}
Working in the hybrid framework of the high energy $pA$ collisions we identify a new contribution to transverse single spin asymmetry (SSA). The phase necessary for the SSA is provided by the Pomeron-Odderon interference in the dense nuclear target. The complete formula for the $pA \to h X$ polarized cross section also contains the transversity distribution for the polarized projectile as well as the real part of the twist-3 fragmentation function. We numerically estimate the asymmetry $A_N$ and its nuclear dependence. Based on a model computation we find that $A_N$ can be a percent level in the forward and low-$P_{h\perp}$ region. For large nuclei we find significant suppression, with $A_N \propto A^{-7/6}$ parametrically. As a notable feature we find a node of $A_N$ as a function of the $P_{h\perp}$ around the values of the initial saturation scale that could be used to test this mechanism experimentally.
\end{abstract}

\maketitle

\section{Introduction and motivation}

Single spin asymmetry (SSA) is left-right asymmetric particle production in hadronic collisions involving a transversely polarized proton, determined through a ratio
\be
A_N = \frac{1}{2}\frac{\rmd \sigma(\uparrow) - \rmd \sigma(\downarrow)}{\rmd \sigma}=\frac{1}{2}\frac{\rmd\Delta\sigma}{\rmd\sigma}\,,
\label{eqn:an}
\ee
where ($\rmd \sigma$) $\rmd\Delta\sigma$ is the (un)polarized cross section and in the numerator the difference between opposite transverse spins of the polarized proton is taken.  Notwithstanding the common folklore of vanishing $A_N$ at high energies \cite{Kane:1978nd}, experiments revealed that $A_N$ is by contrast very large - at the top RHIC energies $A_N$ reaches up to $\sim 10 \%$ in the forward region of the produced hadron
\cite{STAR:2003lxu,PHENIX:2005jxc,STAR:2008ixi,BRAHMS:2008doi,STAR:2012hth,PHENIX:2014qwb,PHENIX:2019ouo,PHENIX:2023xnx,PHENIX:2023axd}. The fundamental origin of large SSA is one of the main open questions in spin physics.

The theoretical description of SSA \cite{Barone:2001sp,DAlesio:2007bjf,Pitonyak:2016hqh} in $pp$ collisions relies on uncovering the mechanism that is sensitive to the phase in the amplitude interferences based on which $\rmd\Delta\sigma$ is computed. In the collinear twist-3 framework the known mechanisms involve the twist-3 distribution functions \cite{Efremov:1981sh,Efremov:1984ip,Sivers:1989cc,Sivers:1990fh,Qiu:1991pp,Qiu:1991wg} (also from the unpolarized hadron \cite{Kanazawa:2000hz}) whereby the phase is picked up from the imaginary parts of the scattering kernels \cite{Kouvaris:2006zy,Ji:2006ub,Eguchi:2006qz,Eguchi:2006mc}. Another possibility relies on the imaginary part of the chiral-odd twist-3 fragmentation functions (FF) \cite{Collins:1992kk,Anselmino:1994tv,Yuan:2009dw,Kang:2010zzb,Metz:2012ct,Kanazawa:2013uia}.

Large SSAs in the forward region sparked an interest in the small-$x$ community as the unpolarized target naturally resides in the regime of large densities, dominantly populated by gluons. This lead to a surge in activities on the crossroads between transverse spin and small-$x$ physics within the $k_\perp$-factorization approach \cite{Boer:2006rj,Kang:2011ni,Kang:2012vm,Schafer:2013wca,Altinoluk:2014oxa,Schafer:2014zea,Schafer:2014xpa,Boer:2015pni,Yao:2018vcg}, the hybrid approach \cite{Kovchegov:2012ga,Hatta:2016wjz,Hatta:2016khv,Benic:2018amn,Kovchegov:2020kxg,Benic:2022qzv} and also computations of polarized transverse momentum distributions \cite{Zhou:2013gsa,Kovchegov:2013cva,Dong:2018wsp,Boussarie:2019vmk,Kovchegov:2021iyc,Boer:2022njw,Kovchegov:2022kyy,Santiago:2023rfl,Zhu:2024iwa}. SSA in $pA$ collisions have recently been measured by the PHENIX \cite{PHENIX:2019ouo,PHENIX:2023xnx,PHENIX:2023axd} and the STAR \cite{STAR:2020grs} collaborations at RHIC. Fitting the nuclear dependence of SSA as $A_N\propto A^P$, the PHENIX collaboration found $P = - 0.37$, while STAR collaboration reports a weaker $A$ dependence $P = -0.027 \pm 0.005$, albeit in a range of larger $x_F$ (Feynman-$x$). In Ref.~\cite{Benic:2018amn} one of us, together with Y. Hatta, confronted the twist-3 FF contribution to SSA based on the hybrid framework \cite{Hatta:2016khv} to the RHIC data. It was realized that the parametric $A$-dependence of $A_N$ set by the initial condition $A_N \propto A^{-1/3}$ \cite{Hatta:2016khv} is washed away by the small-$x$ evolution. This result was found to be in line with the STAR data, but the stronger nuclear suppression found by the PHENIX collaboration could not be explained.

In this work we bring to attention a new mechanism for the phase required for the SSA, and first pointed out in the work by Y.~Kovchegov and M.~Sievert \cite{Kovchegov:2012ga}. The scenario in \cite{Kovchegov:2012ga} relies on the small-$x$, or Color Glass Condensate (CGC) approach \cite{Kovchegov:2012mbw,Iancu:2003xm,Gelis:2010nm,Blaizot:2016qgz} describing the collision through all-twist eikonal interactions with the target whereby the underlying target gluon distribution is in principle a complex function, with its real and imaginary part constituting the Pomeron and the Odderon. By considering scattering of polarized quark on a dense target Ref.~\cite{Kovchegov:2012ga} provided a proof-of-principle demonstration through which the phase required for the SSA can be realized by the Pomeron-Odderon interferences. We take a step further by embedding this idea in a hadronic collision. The distribution of transversely polarized quark inside the transversely polarized proton is described through the chiral-odd collinear transversity distribution. The description of the produced unpolarized hadron relies on chiral-odd (and in general complex) twist-3 FFs. In Ref.~\cite{Hatta:2016khv} only the imaginary part of the twist-3 FFs was considered, and the result was proportional to the Pomeron distribution in the target. We find that picking up the real part of the twist-3 FF, there is a new contribution proportional to the Pomeron-Odderon interferences.

The magnitude and the functional dependence of the Odderon exchange as well as the twist-3 FFs plays a key role in determining the overall size and the functional form of $A_N$. 
Our model estimate, using a recent microscopic computations for the Odderon exchange \cite{Dumitru:2018vpr,Dumitru:2019qec,Dumitru:2020fdh,Dumitru:2022ooz,Benic:2023ybl}, and a chiral quark model for the real part of the twist-3 FF \cite{Ji:1993qx,Yuan:2003gu}, leads to about a percent level $A_N$ in $pp$ collisions at large $x_F \gtrsim 0.6$ and produced hadron momenta $P_{h\perp} \sim Q_S$, where $Q_S$ is the initial saturation scale. For $P_{h\perp}$ of about a few GeV, the polarized cross section has a strong fall-off and $A_N$ drops to per-milles. For nuclear targets we find a strong suppression $A_N \propto A^{-7/6}$ parametrically that pertains also after small-$x$ evolution of the unpolarized target. This leads to a per-mille $A_N$ in $pA$ collisions in a broad kinematic range, making it difficult to explain the current experimental data based solely on Pomeron-Odderon interference.
As an interesting feature, we find that the small-$x$ Odderon exchange leads to a sign change of $A_N$ as a function of $P_{h\perp}$ for $P_{h\perp} \sim Q_S$. The resulting node position of $A_N$ has a very weak $x_F$-dependence, that shifts to higher $P_{h\perp}$ for heavier targets. According to the current data, $A_N$ has a fixed sign across all measured kinematics. A measurement of $A_N$ in the lower $P_{h\perp}$ region can be used as a test of the Pomeron-Odderon interference contribution.

In order to set our notations, in Sec.~\ref{sec:xsect} we quickly remind on the the unpolarized cross section $\rmd\sigma$ in high energy forward $pA$ collisions. The rest of Sec.~\ref{sec:xsect} is devoted to a derivation of the contribution to $\rmd \Delta\sigma$ from the Pomeron-Odderon interference and leading to the main formula given in \eqref{eqn:CSPOfinal}. In the following Sec.~\ref{sec:num} we numerically estimate $A_N$ based on a model computation. In the final Sec.~\ref{sec:concl} we make our conclusions.

\section{Computation of the cross section}
\label{sec:xsect}
Our computation relies on the hybrid approach \cite{Dumitru:2002qt,Dumitru:2005gt} which is appropriate for forward production in hadronic collisions. For the unpolarized $p(P) A(P') \to h(P_h) X$ cross section, the familiar leading order forward formula reads
\be
\frac{\rmd\sigma}{\rmd y_h\rmd^2\Phpv}=\int_{z_{\rm min}}^1\frac{\rmd z}{z^2}D(z)x_qf(x_q)F(x_g,P_{h\perp}/z)\,.
\label{eqn:unpl}
\ee
$f(x_q)$ is the collinear unpolarized parton distribution function (PDF) with $x_q$ being the longitudinal momentum fraction carried by the interacting quark in the projectile with momenta $p^\mu = x_q P^\mu$. $D(z)$ is the unpolarized fragmentation function. 
In the hybrid framework, the target is considered to all twists with eikonal scattering on the target gluon field $A_a^-(x)$ embedded in the Wilson line $V(\xp)=\mathcal{P}\exp[\rmi g\int_{-\infty}^{+\infty}\rmd x^+A_a^-(x^+,0,\xp)t^a]$ (we are working in the covariant gauge where the leading component of the gluon field is $A_a^-(x)$ as sourced by the eikonal current of the target $J_a^{-}(x)$ \cite{Blaizot:2004wu}). The target distribution $F(x_g,k_\perp)$ is the real part of the forward dipole distribution \cite{Dominguez:2011wm} defined as
\be
F(x_g,k_\perp)\equiv \frac{1}{(2\pi)^2}\int_{\xp\yp}\mathrm{e}^{\rmi\kp\cdot(\xp - \yp)}{\rm Re}\langle\calD(\xp,\yp)\rangle_Y\,,
\label{eqn:POMF}
\ee
with the dipole correlator $\calD(\xp,\yp)$ given by
\be
\calD(\xp,\yp) = \frac{1}{N_c}{\rm tr}_c\left(V(\xp)V^\dag(\yp)\right)\,,
\label{eqn:dipole}
\ee
where the gluon momenta is given as $k^\mu = (0,x_gP'^-,\kp)$. We have $\langle \dots \rangle_Y$ as a shorthand for $\frac{\langle P' |\dots|P'\rangle} {\langle P'|P'\rangle}$ and $Y$ is the target rapidity supplied by the Balitsky-Kovchegov (BK) evolution equation \cite{Balitsky:1995ub,Kovchegov:1999yj,Balitsky:2006wa} in the large-$N_c$ limit. $Y$ is related to the momentum fraction $x_g$ as $Y = \log(x_0/x_g)$, where $x_0$ is set by the initial condition for the evolution. 
The projectile (target) momenta fractions $x_q$ ($x_g$) are given as $x_{q,g}=\frac{\Php}{z\sqrt{s}}\mathrm{e}^{\pm y_h}$, where $P_{h_\perp}$ and $y_h$ are the produced hadron transverse momenta and rapidity and $\sqrt{s} = (P+P')^2 = 2 P^+ P'^-$ is the hadronic center-of-mass collision energy. We also have $z_{\rm min}=\frac{\Php}{\sqrt{s}}\mathrm{e}^{y_h}$.

For the polarized cross section we start from an all-order twist-3 formula initially derived in \cite{Kanazawa:2013uia} for semi-inclusive deep inelastic scattering (SIDIS) and subsequently extended to forward $p^\uparrow A\to hX$ collisions in \cite{Hatta:2016khv}. The complete formula reads:
\be
\begin{split}
    \frac{\rmd\Delta\sigma}{\rmd y_h\rmd^2\Phpv}&=\frac{1}{2(2\pi)^3}\Bigg(\int\frac{\rmd z}{z^2}\tr[\Delta(z)S^{(0)}(z)]+\int\frac{\rmd z}{z^2}\mathrm{Im}\tr\left[\Delta^\alpha_\partial(z)\frac{\partial S^{(0)}(K)}{\partial K^\alpha}\right]_{K=\frac{P_h}{z}}\\
    &-\int\frac{\rmd z \rmd z_1}{z^2 z_1^2}P\left(\frac{1}{\frac{1}{z}-\frac{1}{z_1}}\right)\mathrm{Im}\tr[\Delta_F^\alpha(z_1,z)S_\alpha^{(1)L}(z_1,z)+\bar{\Delta}_F^{\alpha}(z,z_1)S_\alpha^{(1)R}(z_1,z)]\Bigg)\,,
\end{split}
\label{polCS}
\ee 
The three components of \eqref{polCS} are commonly referred to as \textit{intrinsic}, \textit{kinematical} and \textit{dynamical} contributions, respectively. The quantities $\Delta(z)$, $\Delta^\alpha_\pd(z)$ and $\Delta_{F}^\alpha(z_1,z)$ represent fragmentation correlators, while $S^{(0)}(z)$ and $S_\alpha^{(1)L,R}(z_1,z)$ are scattering kernels, containing parton correlators from the projectile and the target. We have \cite{Kanazawa:2013uia}: $S_\alpha^{(1)R}(z_1,z) \equiv \gamma^0S_\alpha^{(1)L\dagger}(z,z_1)\gamma^0$, while a bar over a matrix $\Gamma$ in Dirac space is a shorthand for $\bar{\Gamma} \equiv \gamma^0 \Gamma^\dag \gamma^0$. 

The twist-3 FFs contained in the $\Delta$ correlators are defined as follows\footnote{We use the following conventions: $\gamma_5=\rmi\gamma^0\gamma^1\gamma^2\gamma^3$ and $\epsilon_{0123}=+1$.}\cite{Ji:1993vw,Kanazawa:2013uia,Hatta:2016khv}:
\be
\begin{split}
    &\Delta(z) =\frac{M_N}{z} \hat{e}_1(z) + \frac{M_N}{2z}\sigma_{\lambda\alpha} \rmi\gamma_5\epsilon^{\lambda\alpha w P_h}\hat{e}_{\Bar{1}}(z)+\dots\\
    &\Delta_\partial^\alpha(z)=\frac{M_N}{2}\gamma_5\slashed{q}\gamma_\lambda\epsilon^{\lambda\alpha w P_h}\tilde{e}(z)+\dots\\
    &\Delta_F^\alpha(z_1,z)=\frac{M_N}{2}\gamma_5\slashed{q}\gamma_\lambda\epsilon^{\lambda\alpha w P_h}\hat{E}_F(z_1,z)+\dots\,.
\end{split}
\label{eqn:corel}
\ee
where we explicitly kept only the terms that will be important for our calculation - the complete expressions can be found in \cite{Kanazawa:2013uia}. In the above formulas $M_N$ stands for the nucleon mass, while the four-vector $w^\mu$, given by 
$w^\mu=(P_h^-,P_h^+,-\Phpv)/(2E_h^2)\approx (P_h^-,P_h^+,-\Phpv)/(P_h^+)^2$, satisfies $w^2=0$ and $P_h\cdot w=1$. Here $q^\mu=\frac{P_h^\mu}{z}$ is the momenta of the fragmenting quark. The intrinsic twist-3 FFs $\hat{e}_1(z)$ and $\hat{e}_{\bar{1}}(z)$ are real functions, the kinematic function $\tilde{e}(z)$ is purely imaginary, and the dynamical function $\hat{E}_F(z_1,z)$ is in general a complex function. These FFs are not independent. In particular, they obey the relations derived from the QCD equation of motion \cite{Metz:2012ct}, which, after separating the real and imaginary parts, reads \cite{Kanazawa:2015ajw,Koike:2015yza}
\be
\begin{split}
    & z\int_z^\infty\frac{\rmd z_1}{z_1^2}P\left(\frac{1}{\frac{1}{z_1}-\frac{1}{z}}\right)\mathrm{Re}\hat{E}_F(z_1,z) = \hat{e}_1(z)\,,\\
    & z\int_z^\infty\frac{\rmd z_1}{z_1^2}P\left(\frac{1}{\frac{1}{z_1}-\frac{1}{z}}\right)\mathrm{Im}\hat{E}_F(z_1,z) + z\mathrm{Im}\tilde{e}(z) = \hat{e}_{\bar{1}}(z)\,.
\end{split}
\label{eqn:QCD_of_motion} 
\ee

Since the twist-3 FFs in the intrinsic and kinematical parts of \eqref{polCS} are purely real or purely imaginary, respectively, the scattering kernel $S^{(0)}(z)$, and its non-collinear extension $S^{(0)}(K)$, both have to be real. By contrast, the dynamical twist-3 FFs are complex and therefore in principle both the real and imaginary parts of the scattering kernel $S_{L,R}^{(1)\alpha}(z_1,z)$ can contribute. More explicitly, we can rewrite the dynamical twist-3 part of \eqref{polCS} as
\be
\begin{split}
    \frac{\rmd\Delta\sigma_{\rm dyn-tw3}}{\rmd y_h\rmd^2\Phpv}&=-\frac{1}{2(2\pi)^3}\frac{M_N}{2}\epsilon^{\lambda\alpha w P_h}\int\frac{\rmd z_1 \rmd z}{z_1^2z^2}P\left(\frac{1}{\frac{1}{z}-\frac{1}{z_1}}\right)\\
    &\times\Bigg(\mathrm{Re} \hat{E}_F(z_1,z)\times\mathrm{Im}\tr\left(\gamma_5\slashed{q}\gamma_\lambda S_\alpha^{(1)L}(z_1,z)+\gamma_\lambda\slashed{q}\gamma_5\Bar{S}_\alpha^{(1)L}(z_1,z)\right)\\
    &+\mathrm{Im} \hat{E}_F(z_1,z)\times\mathrm{Re}\tr\left(\gamma_5\slashed{q}\gamma_\lambda S_\alpha^{(1)L}(z_1,z)-\gamma_\lambda\slashed{q}\gamma_5\Bar{S}_\alpha^{(1)L}(z_1,z)\right)\Bigg)\,.
\end{split}
\label{eqn:twocontr}
\ee
where we have used $z_1\leftrightarrow z$ for the second term in the third line of \eqref{polCS}. The difference in the relative sign between the second and the third line in \eqref{eqn:twocontr} comes from the hermitian conjugation of the $\Delta_F^\alpha(z_1,z)$ correlator. In the third line, proportional to the imaginary part of the dynamical twist-3 FF, the scattering kernel is real. This is a common source of {\it single} spin asymmetry\footnote{To access the real parts of the fragmentation functions one considers instead {\it double} spin asymmetry, see e.~g.~\cite{Jaffe:1993xb,Yuan:2003gu,Kanazawa:2014tda,Koike:2015yza}.} in the collinear framework \cite{Metz:2012ct,Kanazawa:2013uia} and also in the hybrid framework \cite{Hatta:2016khv}. For the second line to contribute to SSA we need imaginary scattering kernel that is possible at higher orders from loop diagrams \cite{Gamberg:2018fwy,Benic:2019zvg,Kovchegov:2020kxg}. 

Another way to generate SSA, as proposed in \cite{Kovchegov:2012ga}, is through higher twist corrections in the {\it unpolarized} target, which is the mechanism we pursue in this work. In general, the off-forward dipole correlator \eqref{eqn:dipole} can be decomposed in its real and imaginary part as
\be
\left<\calD(\xp,\xp')\right>_Y=\calP_Y(\xp,\xp')+\rmi\calO_Y(\xp,\xp')\,.
\label{eqn:PO}
\ee 
The real part represents the Pomeron exchange, given by two gluons at the lowest order in the expansion of the Wilson line. In a simple $q-q$ scattering process, the Pomeron is a vertical cut across the quark currents of the two-gluon ladder diagram \cite{Forshaw:1997dc}. Likewise, the Odderon is given as a three-gluon ladder with the $d^{abc}$ color factor and with two vertical cuts, appropriate for eikonal currents at high energies. Indeed, the lowest order contribution to the Odderon exchange from \eqref{eqn:PO} is given by three gluons with the symmetric color factor $d^{abc}$ \cite{Ewerz:2003xi,Hatta:2005as} that emerges in an off-forward amplitude. Pomeron and Odderon can also be understood in terms of transverse momentum distributions (GTMDs). The correlator in \eqref{eqn:PO} is related to a dipole-type GTMD at small-$x$ \cite{Dominguez:2011wm,Boer:2018vdi} as
\be
\begin{split}
\frac{2}{P'^-}&\int \frac{\rmd z^- \rmd^2 \zp}{(2\pi)^3} \rme^{\rmi x P'^- z^+ - \rmi \kp \cdot \zp} \langle P_2|{\rm tr}[F^{+i}(-z/2) W^\dag F^{+j} (z/2) W']| P_1\rangle\\
& \to \frac{2 N_c}{ (2\pi)^4 \alpha_S} \left(k_\perp^i + \frac{\Delta_\perp^{i}}{2} \right)\left(k_\perp^j - \frac{\Delta_\perp^{j}}{2} \right)\int_{\xp\yp} \rme^{-\rmi \kp \cdot(\xp - \yp)} \rme^{+\rmi \Dp \cdot (\xp + \yp)/2} \frac{\langle P_2|\calD(\xp,\yp)| P_1\rangle}{\langle P'|P' \rangle}\,,
\end{split}
\label{eq:gtmd}
\ee
where $P_{1,2} = P' \mp \Delta/2$ and $W$, $W'$ symbolically represent Wilson lines required for gauge invariance. GTMDs are in general complex functions \cite{Meissner:2009ww,Lorce:2013pza}, and so \eqref{eq:gtmd} shows that its real and imaginary parts are controlled by the Pomeron and the Odderon exchanges at small-$x$.

\begin{figure}[htb]
\includegraphics[scale=2.2]{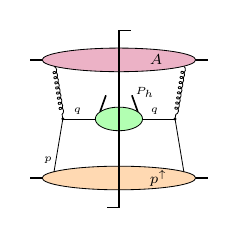}
\includegraphics[scale=2.2]{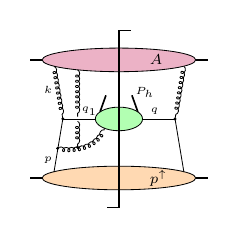}
\caption{Diagrams contributing to polarized cross section in the leading order. Left diagram corresponds to the two-body contribution $S^{(0)}(z)$, while the right diagram corresponds to the three-body contribution $S_\alpha^{(1)L}(z_1,z)$ from \eqref{polCS}. The bottom blob represents the polarized proton, as given by the transversity distribution. The middle blob represents the fragmentation process, and the upper blob represents the unpolarized target. The vertical gluons stand for all-order eikonal interactions with the target through the Wilson lines. Only the right diagram gives a contribution from the Odderon exchange to SSA.}
\label{fig:diagcom}
\end{figure}

We now work out the relationship between the Wilson line correlators and the scattering kernel. We are focusing on  $S^{(1)}_{L,\alpha}(z_1,z)$, i.e. only on the right diagram from Fig. \ref{fig:diagcom}, because this is the only source of the Odderon contribution \cite{Kovchegov:2012ga,Benic:2022qzv}. The scattering kernel $S^{(1)}_{R,\alpha}(z_1,z)$ is given by the mirror image of this diagram. An explicit expression for $S^{(1)}_{L,\alpha}(z_1,z)$ reads
\be
S_L^{(1)\alpha}(z_1,z)=\frac{1}{2P^+}\frac{2}{N_c^2-1}\int\rmd x_q (2\pi)\delta\left(q^+-p^+\right)\mathrm{tr}_c\left\langle\calM^{\alpha,a} \Phi(x_q)\bar{\mathcal{M}}t^a\right\rangle_Y\,,
\label{eqn:Sl}
\ee
where the color factor $2 t^a/(N_c^2 - 1)$ is associated with the fragmentation function in color space, that is $\Delta_F^{\alpha,a}(z_1,z) = 2 t^a \Delta_F^\alpha(z_1,z)/(N_c^2 - 1)$ fixed such that ${\rm tr}_c(t^a \Delta_F^{\alpha,a}) = \Delta_F^\alpha$. $\Phi(x_q)$ is the collinear quark correlator and we pick up the projection onto the twist-2 transversity distribution $h_1(x_q)$ as
\be
\Phi(x_q) = -\frac{\rmi}{2}\gamma_5 P^+ S_{\perp i}\sigma^{-i} h_1(x_q) + \dots\,.
\ee
The amplitudes $\calM_\alpha^a$ and $\calM$ were already found in \cite{Benic:2022qzv} for different purpose. The explicit expressions are\footnote{In Ref.~\cite{Benic:2022qzv} we missed a factor of $-\rmi$ in defining $\calM$.} 
\be
\begin{split}
&\mathcal{M}_\alpha^a = \rmi\int_{\kp} \int_{\xp\yp}e^{\rmi\kp\cdot\xp}e^{\rmi\left(\qp-\kp\right)\cdot\yp}T_\alpha(\kp)V(\xp)t^bU^{ab}(\yp)\,,\\
&\mathcal{M} =-\rmi\int_{\kp}\int_{\xp\yp}e^{\rmi\kp\cdot\xp}e^{\rmi\left(\qp-\kp\right)\cdot\yp}\gamma^+V(\xp)\,,
\end{split}
\label{eqn:amplitudes}
\ee
where $U^{ab}(\yp)$ is the adjoint Wilson line and
\be
T_\alpha(\kp)=\rmi\int_{-\infty}^{+\infty}\frac{\rmd k^-}{(2\pi)}\gamma^+\frac{\slashed{q}_1-\slashed{k}}{\left(q_1-k\right)^2+\rmi\epsilon}\gamma^\beta\frac{V_{\alpha\beta}}{\left(p + k-q_1\right)^2+\rmi\epsilon}\,.
\label{eqn:tensor}
\ee
Following \cite{Hatta:2016khv} we are working in the covariant gauge \cite{Blaizot:2004wu}, taking the full triple-gluon vertex $V_{\alpha\beta}$
\be
V_{\alpha\beta}=\delta_\beta^+\left(q_1-q-2p - 2k\right)_\alpha+2g_{\alpha\beta}\left(q-q_1\right)^++\delta_\alpha^+\left(q_1+p+k\right)_\beta\,.
\label{eqn:vertex}
\ee
Here $k^\mu$ is the gluon momenta from the target attaching onto the quark line, see Fig.~\ref{fig:diagcom}, while $q_1^\mu = P_h^\mu/z_1$ and $q^\mu = P_h^\mu/z$ are final quark momenta from opposite sides of the final state cut. In all the above formulas we use the notations $\int_{\kp}\equiv\int\frac{\rmd^2\kp}{(2\pi)^2}$ and $\int_{\xp}\equiv\int\rmd^2\xp$ for transverse momentum and transverse coordinate integrals, respectively.

We now calculate the trace appearing in \eqref{eqn:twocontr}. Inserting \eqref{eqn:Sl} into \eqref{eqn:twocontr} we have
\be
\begin{split}
    &\epsilon^{\lambda\alpha w P_h}{\rm tr}\left(\gamma_5\slashed{q}\gamma_\lambda S_{L,\alpha}^{(1)}(z_1,z)\right)=\\
    &=\frac{1}{2(N_c^2-1)}\int\rmd x_q h_1(x_q)(2\pi)\delta\left(q^+-p^+\right)\int_{\xp\yp\xp'\yp'}\int_{\kp\kp'}\rme^{\rmi\kp\cdot\xp}\rme^{\rmi(\qp-\kp)\cdot\yp}\rme^{-\rmi\kp'\cdot\xp'}\rme^{-\rmi\left(\qp-\kp'\right)\cdot\yp'}\\
    &\epsilon^{\lambda\alpha w P_h} S_{\perp i}\mathrm{tr}\left[\rmi\gamma_5\sigma^{-i}\gamma^+\gamma_5\slashed{q}\gamma_\lambda T_\alpha(\kp)\right]{\rm tr}_c\left\langle V^\dagger(\xp')t^b V(\xp)t^a U^{ba}(\yp)\right\rangle_Y\,.
\end{split}
\label{eqn:LONG}
\ee
The remaining Dirac trace reads
\be
\begin{split}
    &S_{\perp i}\epsilon^{\lambda\alpha w P_h}\mathrm{tr}\left[\rmi\gamma_5\sigma^{-i}\gamma^+\gamma_5\slashed{q}\gamma_\lambda T_{\alpha}(\kp)\right]=\\
    &=\rmi\int_{-\infty}^{+\infty}\frac{\rmd k^-}{(2\pi)}\frac{1}{\left(q_1-k\right)^2+\rmi\epsilon}\frac{V_{\alpha\beta}}{\left(p + k-q_1\right)^2+\rmi\epsilon}S_{\perp i}\epsilon^{\lambda\alpha w P_h}\mathrm{tr}\left[\rmi\gamma_5\sigma^{-i}\gamma^+\gamma_5\slashed{q}\gamma_\lambda \gamma^+\left(\slashed{q}_1 -\slashed{k}\right)\gamma^\beta\right]\,,
    \label{eqn:kmint}
\end{split}
\ee
and in the forward limit we get
\be
S_{\perp i}\epsilon^{\lambda\alpha w P_h}\mathrm{tr}\left[\rmi\gamma_5\sigma^{-i}\gamma^+\gamma_5\slashed{q}\gamma_\lambda \gamma^+\left(\slashed{q}_1 -\slashed{k}\right)\gamma^\beta\right]V_{\alpha\beta}\to 32 q^+ q_1^+\left(\qonp-\kp\right)\times \Sp\,,
\label{eqn:solvedtrace}
\ee
confirming \cite{Hatta:2016khv}. We have abbreviated $\aap\times \bp \equiv -\epsilon^{+- i j} a_\perp^i b_\perp^j$. The support of the twist-3 fragmentation function $\hat{E}_F(z_1,z)$ is given by $0<z<1$, $z<z_1<\infty$ \cite{Meissner:2008yf,Kanazawa:2015ajw} and consequently, the poles for the $k^-$ integration in Eq.~\eqref{eqn:kmint} are necessarily located on opposite sides of the real axis. After performing the $k^-$ integration we find
\be
S_{\perp i}\epsilon^{\lambda\alpha w P_h}\mathrm{tr}\left[\rmi\gamma_5\sigma^{-i}\gamma^+\gamma_5\slashed{q}\gamma_\lambda T_{\alpha}(\kp)\right]=-16q_1^+\frac{(\qonp - \kp)\times \Sp}{\left(\qonp-\kp\right)^2}\,.
\ee
Using $U^{ab}(\xp) = 2{\rm tr}_c \left(t^a V(\xp)t^b V^\dag(\xp)\right)$, the color correlator appearing in \eqref{eqn:LONG} is computed as
\be
\mathrm{tr}_c\left<V^\dagger(\xp')t^bV(\xp)t^a U^{ba}(\yp)\right>_Y=\frac{1}{2}\left(N_c^2\left<\calD(\yp,\xp')\calD(\xp,\yp)\right>_Y-\left<\calD(\xp,\xp')\right>_Y\right)\,.
\ee
Eq.~\eqref{eqn:LONG} can now be written as
\be
\begin{split}
    \epsilon^{\lambda\alpha w P_h}\mathrm{tr}\left[\gamma_5 \slashed{q}\gamma_\lambda S_{L,\alpha}^{(1)}(z_1,z)\right] & =-\frac{8\pi}{N_c^2-1}x_qh_1(x_q)\frac{z}{z_1}\int_{\kp}\int_{\xp\xp'\yp}\mathrm{e}^{\rmi\kp\cdot(\xp-\yp)}\mathrm{e}^{\rmi \qp\cdot(\yp-\xp')}\\
    &\times\frac{(\qonp - \kp)\times \Sp}{\left(\qonp -\kp\right)^2}\left(N_c^2\left<\calD(\yp,\xp')\calD(\xp,\yp)\right>_Y-\left<\calD(\xp,\xp')\right>_Y\right)\,.
    \label{eqn:firsttrace}
\end{split}
\ee
We have performed the integration over the $\delta$-function in \eqref{eqn:LONG} so that $x_q = q^+/P^+$.
As for the second trace in Eq.~\eqref{eqn:twocontr}, we have
\be
    \epsilon^{\lambda\alpha w P_h}\mathrm{tr}\left(\gamma_\lambda\slashed{q}\gamma_5\Bar{S}_{L,\alpha}^{(1)}(z_1,z)\right)
=-\epsilon^{\lambda\alpha w P_h}\mathrm{tr}\left(\gamma_5\slashed{q}\gamma_\lambda S_{L,\alpha}^{(1)}(z_1,z)\right)^*\,.
\label{eq:2ndtr}
\ee
Plugging \eqref{eqn:firsttrace} and \eqref{eq:2ndtr} into \eqref{eqn:twocontr} we find
\be
\begin{split}
\frac{\rmd\Delta\sigma_{\rm dyn-tw3}}{\rmd y_h\rmd^2\Phpv}&=\frac{M_N}{4\pi^2}\frac{N_c^2}{N_c^2-1}\int\frac{\rmd z \rmd z_1}{z^2 z_1^2}P\left(\frac{1}{\frac{1}{z}-\frac{1}{z_1}}\right)x_qh_1(x_q)\int_{\kp}\frac{z}{z_1}\frac{(\qonp - \kp)\times \Sp}{\left(\qonp-\kp\right)^2}\\
&\times\Big\{\mathrm{Re} \hat{E}_F(z_1,z)\mathrm{Im}\int_{\xp\xp'\yp}\rme^{\rmi\kp\cdot(\xp-\yp)}\rme^{\rmi\qp\cdot(\yp-\xp')}\\
&\times\Big(\left<\calD(\yp,\xp')\calD(\xp,\yp)\right>_Y -\left<\calD(\yp,\xp)\calD(\xp',\yp)\right>_Y -\langle\calD(\xp,\xp')\rangle_Y + \langle\calD(\xp',\xp)\rangle_Y\Big)\\
& +\mathrm{Im} \hat{E}_F(z_1,z)\mathrm{Re}\int_{\xp\xp'\yp}\rme^{\rmi\kp\cdot(\xp-\yp)}\rme^{\rmi\qp\cdot(\yp-\xp')}\\
&\times\Big(\left<\calD(\yp,\xp')\calD(\xp,\yp)\right>_Y+\left<\calD(\yp,\xp)\calD(\xp',\yp)\right>_Y -\langle\calD(\xp,\xp')\rangle_Y - \langle\calD(\xp',\xp)\rangle_Y\Big)\Big\}\,,
\end{split}
\ee
where we used $\calD(\xp,\xp')^*=\calD(\xp',\xp)$, and rotational invariance of QCD $\calD(-\xp,-\xp')=\calD(\xp,\xp')$.
The double-dipole contribution can be further decomposed in the large-$N_c$ limit as $\left<\calD\calD\right>_Y \to \left<\calD\right>_Y\left<\calD\right>_Y$. Using \eqref{eqn:PO} we find
\be
\begin{split}
     \frac{\rmd\Delta\sigma_{\rm dyn-tw3}}{\rmd y_h\rmd^2\Phpv}&=\frac{M_N}{2\pi^2}\frac{N_c^2}{N_c^2-1}\int\frac{\rmd z \rmd z_1}{z^2 z_1^2}P\left(\frac{1}{\frac{1}{z}-\frac{1}{z_1}}\right)x_qh_1(x_q)\int_{\kp}\frac{z}{z_1}\frac{(\qonp - \kp)\times \Sp}{\left(\qonp - \kp\right)^2}\\
     &\times\Big\{\mathrm{Re} \hat{E}_F(z_1,z)\int_{\xp\xp'\yp} \rme^{\rmi\kp\cdot(\xp-\yp)}\rme^{\rmi\qp\cdot(\yp-\xp')}\\
     &\times\Big(\calP_Y(\xp,\yp)\calO_Y(\yp,\xp')+\calP_Y(\yp,\xp')\calO_Y(\xp,\yp) - \calO_Y(\xp,\xp')\Big)\\
     &+\mathrm{Im} \hat{E}_F(z_1,z)\int_{\xp\xp'\yp}\mathrm{e}^{\rmi\kp\cdot(\xp-\yp)}\mathrm{e}^{\rmi\qp\cdot(\yp-\xp')}\\
     &\times\Big(\calP_Y(\xp,\yp)\calP_Y(\yp,\xp')-\calO_Y(\xp,\yp)\calO_Y(\yp,\xp')  - \calP_Y(\xp,\xp')\Big)\Big\}\,.
\end{split}
\label{eqn: PO eqn}
\ee
where the sign change for the second term in third line and the second term in the fifth line is coming from the symmetry properties $\calP_Y(\xp,\xp') = \calP_Y(\xp',\xp)$ and $\calO_Y(\xp,\xp') = -\calO_Y(\xp',\xp)$ \cite{Hatta:2005as}. For the same reason also the operations ${\rm Im}(\dots)$ and ${\rm Re}(\dots)$ on the Fourier transforms of Pomeron and Odderon distributions are dropped.

Ref.~\cite{Hatta:2016khv} considered the first term and the third term in the fourth line of \eqref{eqn: PO eqn}. Together with the first line in \eqref{polCS} this leads to the expression for the fragmentation contribution to the polarized cross section found in \cite{Hatta:2016khv}. The remaining terms in \eqref{eqn: PO eqn} are new. The term $\sim\calO_Y$ trivially vanishes after the integration over the impact parameter. This leaves $\calP_Y\calO_Y$ and $\calO_Y\calO_Y$ term. It is generally expected that the Odderon is much smaller than the Pomeron (see for example \cite{Yao:2018vcg}), and so we will focus only on the $\calP_Y\calO_Y$ term.

It is useful to write \eqref{eqn: PO eqn} in momentum space. For this purpose we first introduce double Fourier transforms
\be
\calP_Y(\kp,\Dp)=\int_{\xp\yp}e^{\rmi\kp\cdot(\xp - \yp)}e^{\rmi\Dp\cdot(\xp + \yp)/2}\calP_Y(\xp,\yp) = \int_{\rp\bp}e^{\rmi\kp\cdot\rp}e^{\rmi\Dp\cdot\bp}\calP_Y(\rp,\bp)\,,
\ee
and similar for $\calO_Y(\xp,\yp)$, where $\rp = \xp - \yp$ and $\bp = (\xp + \yp)/2$ are the dipole size and impact parameters, respectively. A simple computation gives
\be
\begin{split}
     &\frac{\rmd\Delta\sigma_{\rm dyn-tw3}}{\rmd y_h\rmd^2\Phpv} = \frac{M_N}{2\pi^2}\frac{N_c^2}{N_c^2-1}\int\frac{\rmd z \rmd z_1}{z^2 z_1^2}P\left(\frac{1}{\frac{1}{z}-\frac{1}{z_1}}\right)x_qh_1(x_q)\int_{\kp\Dp}\frac{z}{z_1}\frac{\left(\qonp - \kp - \frac{1}{2}\Dp\right)\times \Sp}{\left(\qonp - \kp - \frac{1}{2}\Dp \right)^2}\\
     &\times\Bigg\{\mathrm{Re} \hat{E}_F(z_1,z)\Bigg[\calO_Y(\kp,\Dp)\calP_Y\Big(\qp - \frac{1}{2}\Dp,-\Dp\Big) + \calP_Y(\kp,\Dp)\calO_Y\Big(\qp - \frac{1}{2}\Dp,-\Dp\Big)\Bigg]\\
     &+\mathrm{Im} \hat{E}_F(z_1,z)\Bigg[\calP_Y\Big(\qp - \frac{1}{2}\Dp,-\Dp\Big)\calP_Y(\kp,\Dp)  - (2\pi)^2 \delta^{(2)}\Big(\kp + \frac{1}{2}\Dp - \qp\Big) \calP_Y\Big(\kp + \frac{1}{2}\Dp,0\Big)\Bigg]\Bigg\}\,.
\end{split}
\label{eqn: PO eqn2}
\ee
where we have shifted $\kp \to \kp + \Dp/2$.
We can intuitively understand the momentum transfer $\Dp$ integration arising from a formation of an intermediate target state on the left side of the cut on Fig.~\ref{fig:diagcom}. The structure is similar to the non-linear part of the BK equation in momentum space, see e.~g.~\cite{Hatta:2022bxn}. Since this is an inclusive process the intermediate momentum transfer $\Dp$ balances out in the Pomeron-Odderon interferences. The intrinsic momenta $\qp -\Dp/2$ and $\kp$ emerge as the momenta difference of the respective (fundamental) Wilson lines coming out of the target. In the first ($\sim \calO_Y \calP_Y$) term in the second line of \eqref{eqn: PO eqn2} the projectile and the target exchange the Odderon in the amplitude alone. The resulting intermediate target state forms the Pomeron out of the Wilson lines in the amplitude and its complex conjugate (after crossing the final state cut). In the second ($\sim \calP_Y \calO_Y$) term the projectile and the target exchange the Pomeron in the amplitude, while the Odderon exchange forms by crossing the final state cut and in combination with the Wilson line in the complex conjugate amplitude.

\subsection{\texorpdfstring{Reproducing the double Pomeron contribution from Ref.~\cite{Hatta:2016khv}}{Reproducing the double Pomeron contribution}}
\label{sec:Hatta}

Before calculating the contribution from Pomeron-Odderon interference, we quickly confirm that the double Pomeron contribution (first term in the second line of \eqref{eqn: PO eqn2}) reproduces the corresponding term in \cite{Hatta:2016khv}. We assume that for a large target, the general distribution $\calP_Y(\rp,\bp)$ approximately factorizes as $\calP_Y(\rp,\bp) \approx \calP_Y(r_\perp)T(b_\perp)$, where $T(b_\perp)$ is a profile function normalized to the target area $\int_{\bp} T(b_\perp) = \pi R_A^2$ and $\calP_Y(r_\perp)$ is related to $F(x_g,k_\perp)$  \eqref{eqn:POMF} as
\be
F(x_g,k_\perp) = \frac{\pi R_A^2}{(2\pi)^2}\int_{\rp} \rme^{\rmi \kp\cdot \rp} \calP_Y(r_\perp)\,.
\ee
In momentum space we can write
\be
\calP_Y(\kp,\Dp) \approx \frac{(2\pi)^2}{\pi R_A^2} F(x_g,k_\perp) T(\Delta_\perp)\,,
\label{eq:pomgtmd}
\ee
with $T(\Delta_\perp)$ being the Fourier transform of $T(b_\perp)$. For a large target, the intermediate momentum transfer is typically small $\Dp \ll \qp$, $\kp$ and so we can approximate the $\calP_Y\calP_Y$ term as
\be
\begin{split}
\int_{\Dp} \calP_Y\Big(\qp - \frac{1}{2}\Dp,-\Dp\Big)&\calP_Y(\kp,\Dp) \approx \frac{(2\pi)^4}{(\pi R_A^2)^2}F(x_g,q_\perp)F(x_g,k_\perp)\int_{\Dp}T^2(\Delta_\perp)\\
& = \frac{(2\pi)^4}{(\pi R_A^2)^2}F(x_g,q_\perp)F(x_g,k_\perp)\int_{\bp}T^2(b_\perp) \approx \frac{(2\pi)^4}{\pi R_A^2} F(x_g,q_\perp)F(x_g,k_\perp)\,,
\end{split}
\ee
which holds for a sharp profile $T^2(b_\perp) \approx T(b_\perp)$.
Setting also $\Dp = 0$ in the hard part of \eqref{eqn: PO eqn2} we can compute the angular part of the $\kp$-integral (see App.~\ref{integrallll})
\be
\int_{\kp}\frac{z}{z_1}\frac{\left(\qonp -\kp\right)\times \Sp}{\left(\qonp -\kp\right)^2} F(x_g,k_\perp) =\frac{z}{2\pi}\frac{\Phpv\times \Sp}{\Phpv^2}\int_0^{q_{1\perp}}k_\perp\rmd k_\perp F(x_g,k_\perp)\,,
\label{eqn:Hattaintegral}
\ee
The contribution to the polarized cross section is
\be
\begin{split}
    \frac{\rmd\Delta\sigma_{\calP\calP}}{\rmd y_h\rmd^2\Phpv} &= 4\pi M_N\frac{N_c^2}{N_c^2-1}\frac{\Phpv\times\Sp}{\Phpv^2}\frac{1}{\pi R_A^2}\int_{z_{\rm min}}^1\frac{\rmd z}{z}\int_z^\infty\frac{\rmd z_1}{z_1^2} P\left(\frac{1}{\frac{1}{z}-\frac{1}{z_1}}\right){\rm Im} \hat{E}_F(z_1,z)\\
    &\times x_qh_1(x_q)F\left(x_g,\frac{\Php}{z}\right) \int_0^{\frac{\Php}{z_1}}k_\perp\rmd k_\perp F(x_g,k_\perp)\,,
\end{split}
\label{eqn:Hatta}
\ee
which exactly reproduces the double Pomeron term in Eq.~(46) from \cite{Hatta:2016khv}.
The remaining terms in the expression for the polarized cross section in \cite{Hatta:2016khv} contain the linear pieces in $F(x_g,k_\perp)$ (including its derivative), which originate from the third term in the last line of \eqref{eqn: PO eqn} and also the first two lines in \eqref{polCS}.

\subsection{The contribution from the Pomeron-Odderon interference}

We now focus on the second line in \eqref{eqn: PO eqn2}, describing the contribution to the polarized cross section from Pomeron-Odderon interference. For the Odderon distribution we assume a decomposition of $\rp$ and $\bp$ dependence as
\be
\calO_Y(\rp,\bp) \approx R_A T'(b_\perp)\calO_Y(r_\perp) (\hat{\boldsymbol{r}}_\perp\cdot \hat{\boldsymbol{b}}_\perp)\,,
\label{eqn:odddecomp}
\ee
that is common for a large nuclei \cite{Jeon:2005cf,Kovchegov:2012ga,Lappi:2016gqe,Zhou:2013gsa,Dumitru:2014dra,Boer:2018vdi}.
$\calO_Y(r_\perp)$ is in principle an unknown non-perturbative function, while $T'(b_\perp)$ is the derivative of $T(b_\perp)$. Unlike the Pomeron, Odderon is explicitly off-forward, as signified by the ``directed flow" correlation $\hat{\boldsymbol{r}}_\perp\cdot \hat{\boldsymbol{b}}_\perp = \cos(\phi_r - \phi_b)$ \cite{Dumitru:2015cfa}.
It is convenient to introduce the Fourier transform of $\calO_Y(r_\perp)$ as
\be
\rmi\cos(\phi_k - \phi_0)G(x_g,k_\perp)\equiv\frac{\pi R_A^2}{(2\pi)^2}\int_{\rp} e^{\rmi\kp\cdot\rp}\calO_Y(r_\perp)\cos(\phi_r - \phi_0)\,,
\label{eq:Godd}
\ee
with $\phi_0$ some reference angle. Then
\be
\calO_Y(\kp,\Dp) = \frac{(2\pi)^2}{\pi R_A^2}G(x_g,k_\perp)R_A\Delta_\perp T(\Delta_\perp)\cos(\phi_k-\phi_\Delta)\,.
\label{eq:oddgtmd}
\ee
For the Pomeron distribution we use \eqref{eq:pomgtmd}. As in Sec.~\ref{sec:Hatta} we expand for small $\Dp\ll \qp$, $\kp$. Since Odderon is at least linear in $\Dp$, we expand the hard coefficient to linear order in $\Dp$ to find
\be
\begin{split}
     \frac{\rmd\Delta\sigma}{\rmd y_h\rmd^2\Phpv}& \approx\frac{M_N}{2\pi^2}\frac{N_c^2}{N_c^2-1}\int\frac{\rmd z \rmd z_1}{z^2 z_1^2}P\left(\frac{1}{\frac{1}{z}-\frac{1}{z_1}}\right)x_qh_1(x_q)\mathrm{Re} \hat{E}_F(z_1,z)\\
     &\times\int_{\kp\Dp}\frac{z}{z_1}\left[-\frac{1}{2}\frac{\Dp\times\Sp}{\left(\qonp-\kp\right)^2}+\frac{(\qonp - \kp)\times \Sp}{(\qonp - \kp)^2}\frac{(\qonp - \kp)\cdot\Dp}{(\qonp-\kp)^2}\right]\\
     &\times \left[\calP_Y(\qp,\Dp)\calO_Y\left(\kp,\Dp\right)-\calP_Y\left(\kp,\Dp\right)\calO_Y(\qp,\Dp)\right]\,,
\end{split}
\label{eqn:CSalm}
\ee
where we used $\calO_Y(\qp,-\Dp)=-\calO_Y(\qp,\Dp)$.
Inserting \eqref{eq:pomgtmd} and \eqref{eq:oddgtmd} into \eqref{eqn:CSalm} we can compute the angular integrals - for computational steps see App.~\ref{integrallll}. We highlight that after the angular integrations are performed the contribution $\sim F(x_g,q_\perp) G(x_g,k_\perp)$ (where the Odderon is formed in the amplitude, while the Pomeron involves the amplitude and its complex conjugate, cf.~the discussion below \eqref{eqn: PO eqn2}) vanishes and only the term $\sim F(x_g,k_\perp) G(x_g,q_\perp)$ (Pomeron formed in the amplitude, Odderon in amplitude and its complex conjugate) remains.
Assuming $T(\Delta_\perp) 
= \pi R_A^2 \rme^{-\Delta_\perp^2 R_A^2/4}$, the $\Delta_\perp$ integral yields $\int_0^\infty \Delta_\perp \rmd\Delta_\perp \Delta_\perp^2 T^2(\Delta_\perp) = 2\pi^2$ and the polarized cross section takes a simple form
\be
\begin{split}
    \frac{\rmd\Delta\sigma}{\rmd y_h\rmd^2\Phpv}&=-\pi M_N \frac{N_c^2}{N_c^2-1}\frac{\Phpv\times\Sp}{\Phpv^2}\frac{1}{\Php R_A}\frac{1}{\pi R_A^2}\int_{z_{\rm min}}^1\frac{\rmd z}{z^2} x_q h_1(x_q) G(x_g,\Php/z)\\
    &\times\int_{z}^\infty\frac{\rmd z_1}{z_1^2}P\left(\frac{1}{\frac{1}{z}-\frac{1}{z_1}}\right) z z_1\mathrm{Re} \hat{E}_F(z_1,z) \int_0^{\frac{\Php}{z_1}}k_\perp \rmd k_\perp F(x_g,k_\perp)\,.
\end{split}
\label{eqn:CSPOfinal}
\ee
Eq.~\eqref{eqn:CSPOfinal} is the central result of our work. Odderon represents an off-forward exchange and so it is more common to find Odderon contributing to exclusive and/or diffractive reactions. The interesting point about \eqref{eqn:CSPOfinal} is that, through interference with the Pomeron, Odderon can also appear in SSA, which is an inclusive observable, with the overall target distribution described as a forward matrix element. In a collinear framework this should be described in terms of a twist-5 double distribution and the Pomeron-Odderon interference is seen as an approximation in terms of a product of single distributions that depend on the impact parameter (or momementum transfer).   

Pomeron-Odderon interference also appears in charge asymmetries \cite{Brodsky:1999mz,Hagler:2002nh,Pire:2008xe} and could even contribute to polarized $\Lambda$ production in $ep$ collisions \cite{Koike:2022ddx,Ikarashi:2022yzg,Ikarashi:2022zeo} at high energies. In the CGC framework, the unpolarized cross section is based on the $q\bar{q}$ production amplitude \cite{Marquet:2009ca}. In the polarized cross section we need interference of $q\bar{q}$ and $q\bar{q} g$ amplitudes, as the final state gluon enters the polarized twist-3 FF describing $\Lambda$ productions. The contribution where the unobserved remnant forms a color singlet at the cross section level leads to the same color structure of the target as the process considered here.

Based on \eqref{eqn:CSPOfinal} and \eqref{eqn:unpl}, we can make a parametric estimate of the nuclear dependence for $A_N$. For a dense target we can use GBW-type  \cite{Golec-Biernat:1998zce} parametrization of the Pomeron distribution
\be
F(x_g,k_\perp) \approx \frac{\pi R_A^2}{\pi Q_S^2} \rme^{-k_\perp^2/Q_S^2}\,.
\label{eqn:POMparam}
\ee 
Here $Q_S$ is the saturation scale of the nuclei $Q_S^2 \propto A^{1/3}$. For the Odderon, we can use the Jeon-Venugopalan model \cite{Jeon:2005cf}, which gives \cite{Zhou:2013gsa,Benic:2023ybl}
\be
G(x_g,k_\perp) \approx  \frac{\pi R_A^2}{\pi Q_S^2} \frac{3(N_c^2 - 4)}{16(N_c^2 - 1)^2} \frac{Q_S^3 R_A^3}{\alpha_S^3 A^2} \frac{k_\perp}{Q_S}\left(\frac{k_\perp^2}{Q_S^2} - 2\right)\rme^{-k_\perp^2/Q_S^2}\,.
\label{eqn:JVod}
\ee
Note that $G(x_g,k_\perp)$ has a node as a function of $k_\perp$ that will lead to a sign change of $A_N$ as a function of $P_{h\perp}$.
\be
\begin{split}
&\int_0^{P_{h\perp}/z_1} k_\perp \rmd k_\perp F(x_g,k_\perp) \propto R_A^2 \propto A^{2/3}\,,\\
& G(x_g,P_{h\perp}/z)\propto \frac{R_A^2}{Q_S^2}\frac{Q_S^3 R_A^3}{A^2} \propto A^{-1/6}\,,
\end{split}
\ee
and so $\rmd \Delta\sigma \propto A^{-5/6}$. For the unpolarized cross section \eqref{eqn:unpl} we have $\rmd \sigma \propto A^{1/3}$ leading to $A_N \propto A^{-7/6}$. The same parametric estimate was also obtained in \cite{Kovchegov:2012ga}. By comparison, the nuclear dependence of the double-Pomeron term in \eqref{eqn:Hatta} is $A_N \propto A^{-1/2}$ at $P_{h\perp} \sim Q_S$.
The difference is in part from different $A$-scaling of $F(x_g,k_\perp\sim Q_S) \propto A^{1/3}$ vs. $G(x_g,k_\perp \sim Q_S) \approx A^{-1/6}$ and in part due to the non-trivial integral in $\Delta_\perp$ that effectively brings an extra factor $1/(P_{h\perp} R_A) \propto A^{-1/2}$ in \eqref{eqn:CSPOfinal}.

\section{Numerical estimate}
\label{sec:num}

In this Section we perform a numerical estimate of $A_N$ for $p p$ and for $p A$ collisions. 
For the Pomeron distribution $\calP_Y(r_\perp)$ we use a fit \cite{Salazar:2021mpv} of the numerical solutions of the BK evolution as follows
\be
\calP_Y(r_\perp)=\exp\left[-\frac{1}{4}r_\perp^2 c_0\exp\left(0.3Y\frac{\rme^{-c_1 r_\perp}}{(c_2r_\perp)^{c_3}}\right)A^{1/3}\ln\left(\frac{1}{r_\perp\Lambda}+\rme\right)\right]\,,
\label{eqn:Pom}
\ee
considering $c_0$, $c_1$, $c_2$ and $c_3$ as free fitting parameters, with $\Lambda = 0.2$ GeV. We have checked that the HERA-constrained numerical solutions of the BK equation for $\calP_Y(r_\perp)$ from \cite{Lappi:2013zma} are decently reproduced with $c_0=0.087$ GeV$^2$, $c_1=0.80$ GeV, $c_2=1$ GeV, $c_3=-0.77$. Based on this fit, the initial saturation scale $Q_S(x_0)$ for the proton, obtained from the definition $\calP_{Y = 0}(r_\perp = \sqrt{2}/Q_S(x_0)) = \rme^{-1/2}$ is $Q_S(x_0) = 0.35$ GeV.  The fit \eqref{eqn:Pom} is used to calculate $F(x_g,k_\perp)$ via \eqref{eqn:POMF} that is used in \eqref{eqn:unpl} as well as in \eqref{eqn:CSPOfinal}.

\begin{figure}[ht]
    \includegraphics[scale=0.5]{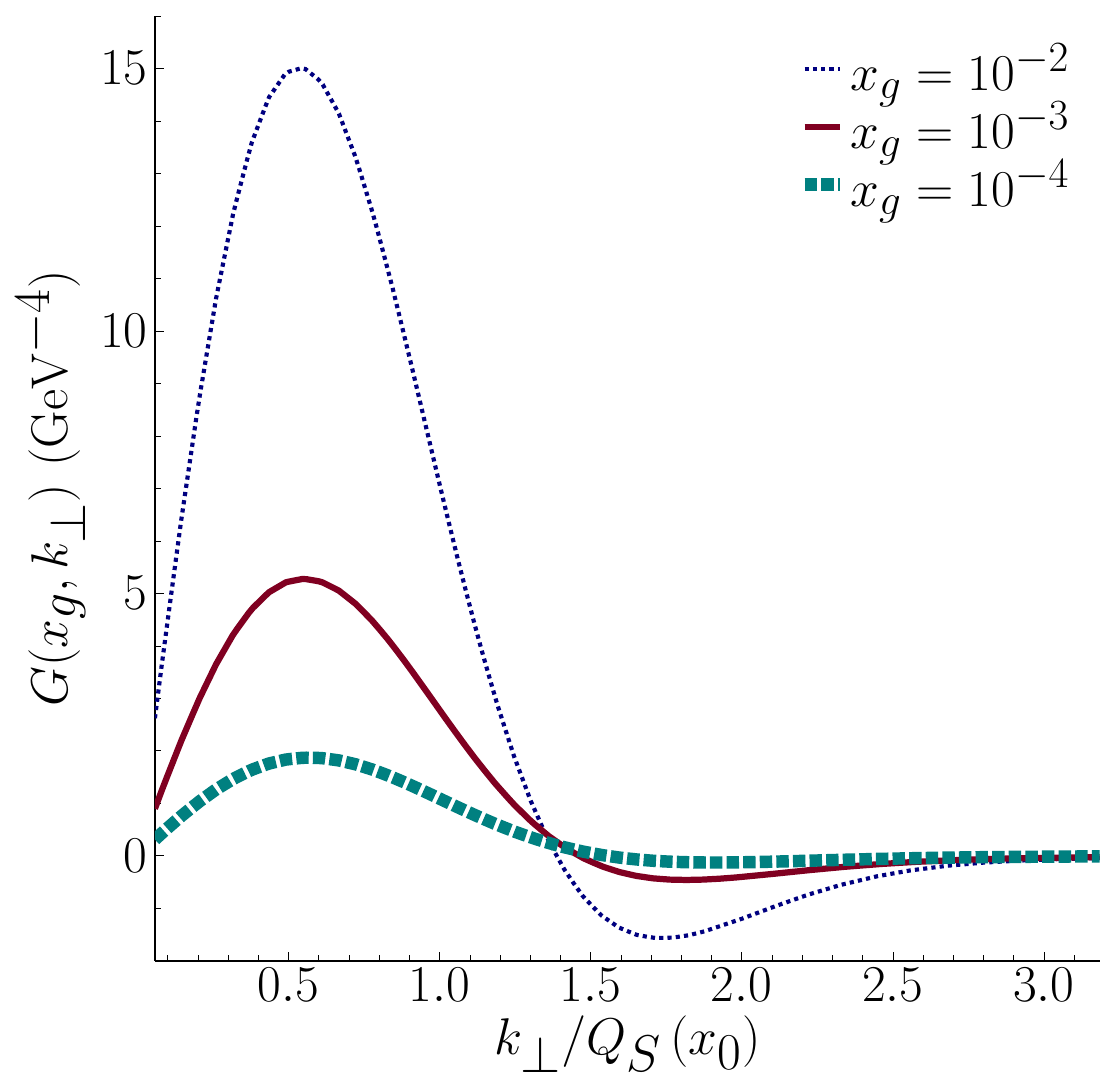}
    \caption{Fourier transform of the Odderon exchange for proton target as a function of $k_\perp$ for different values of $x_g$ as indicated on the plot.}
    \label{fig:FTO}
\end{figure}

For the Odderon we make a similar fit based on the numerical solution \cite{Benic:2023ybl} of its own BK equation \cite{Kovchegov:2003dm,Hatta:2005as,Motyka:2005ep,Lappi:2016gqe,Yao:2018vcg,Contreras:2020lrh}
\be
\calO_Y(r_\perp)=\lambda \mathrm{e}^{- B Y}(c r_\perp)^{3(1+\gamma Y)}\ln\left(\frac{1}{r_\perp\Lambda}+\mathrm{e}\right)\exp\left[-\frac{1}{4}r_\perp^2 c'_0\exp\left (0.15 Y\frac{\rme^{-c_1 r_\perp}}{(c_2r_\perp)^{c_3}}\right)A^{1/3}\ln\left(\frac{1}{r_\perp\Lambda}+\rme\right)\right]\,,
\label{eqn:Odd}
\ee
with free parameters $\lambda$, $B$, $c$, $\gamma$ and $c_0'$. The remaining parameters $c_1$, $c_2$ and $c_3$ are same as in \eqref{eqn:Pom}. For $Y=0$, Eq.~\eqref{eqn:Odd} reduces to the functional form provided by the JV model \cite{Jeon:2005cf,Benic:2023ybl}. We find a reasonable agreement with the numerical solutions for the following parameters values: $B=0.15$, $c=1.21$ GeV, $\gamma=-0.04$ and $c'_0=0.10$ GeV$^{2}$. For the overall $\lambda$ parameter we fix $\lambda = 2.2\times 10^{-4}$ so that the overall magnitude of the Odderon exchange at $x_g  = x_0$ is in line with the recent microscopic computations for proton targets and based on the quark light-front model \cite{Dumitru:2022ooz} which is about an order of magnitude higher than the maximum value set by the group theory constraint \cite{Lappi:2016gqe}. The result for $G(x_g,k_\perp)$ in case of proton target ($A = 1$) is shown on Fig.~\ref{fig:FTO}. The function $G(x_g,k_\perp)$ has a peak at around $k_\perp \approx 0.5 Q_S(x_0)$ and a node around $k_\perp \approx 1.4 Q_S(x_0)$ in this model. The peak position and the node position of $G(x_g,k_\perp)$ does not change with evolution to smaller values of $x_g$, because there is no geometric scaling for the Odderon \cite{Motyka:2005ep,Lappi:2016gqe,Yao:2018vcg}. This seems to be a generic feature of the angular harmonics of the dipole amplitude \cite{Hagiwara:2016kam,Yao:2018vcg}. The main effect of the evolution is a decrease in the magnitude of $G(x_g,k_\perp)$ originating from the non-linear term in the BK equation for the Odderon \cite{Motyka:2005ep,Lappi:2016gqe,Yao:2018vcg}.

Next, we approximate \cite{Hatta:2016khv,Benic:2018amn}
\be
z_1 \int_0^{\frac{\Php}{z_1}} k_\perp \rmd k_\perp F(x_g,k_\perp) \approx z \int_0^{\frac{\Php}{z}}k_\perp \rmd k_\perp F(x_g,k_\perp)\,,
\label{eq:approxz}
\ee
which is to be understood as a zeroth order term\footnote{We numerically confirmed that first order term, proportional to the derivative of \eqref{eq:approxz}, is very small.} in the Taylor series of this function around $z_1 = z$. By using \eqref{eq:approxz}, the remaining $z_1$ dependence in \eqref{eqn:CSPOfinal} is exactly of the form to utilize \eqref{eqn:QCD_of_motion}. This is a great simplification so that the polarized cross section that we will use for the numerical computation becomes
\be
\begin{split}
    \frac{\rmd\Delta\sigma}{\rmd y_h\rmd^2\Phpv}&=-\pi M_N \frac{N_c^2}{N_c^2-1}\frac{\Phpv\times\Sp}{\Phpv^2}\frac{1}{\Php R_A}\frac{1}{\pi R_A^2}\int_{z_{\rm min}}^1\frac{\rmd z}{z^2} z \hat{e}_{1}(z) x_q h_1(x_q) G\left(x_g,\frac{\Php}{z}\right) \int_0^{\frac{\Php}{z}}k_\perp \rmd k_\perp F(x_g,k_\perp)\,,
\end{split}
\label{eqn:csapprox}
\ee
whereby only $\hat{e}_1(z)$ appears! The twist-3 FF $\hat{e}_1(z)$ is typically found to contribute to double spin asymmetry \cite{Jaffe:1993xb,Yuan:2003gu,Kanazawa:2014tda,Koike:2015yza,Bauer:2022mvl}. In the numerical computation we use the result from the chiral quark model \cite{Ji:1993qx,Yuan:2003gu}:
\be
\hat{e}_1(z) = \frac{z}{1-z}\frac{M_q}{M_N} D(z)\,,
\ee
where $M_q$ is the constituent quark mass, approximated as $M_q \approx M_N/3$.

The remaining quantities to determine $A_N$ are as follows. We are using the central tables from the JAM collaboration for the transversity PDF $h_1(x_q)$ \cite{JAMGamberg:2022kdb} as well as for the unpolarized FF $D(z)$ \cite{JAMEthier:2017zbq}, respectively. For unpolarized PDF we are using \cite{Dulat:2015mca}. The factorization scale implicitly present in the collinear PDFs and FFs is set to $P_{h\perp}$. 

\begin{figure}[ht]
        \includegraphics[scale=0.22]{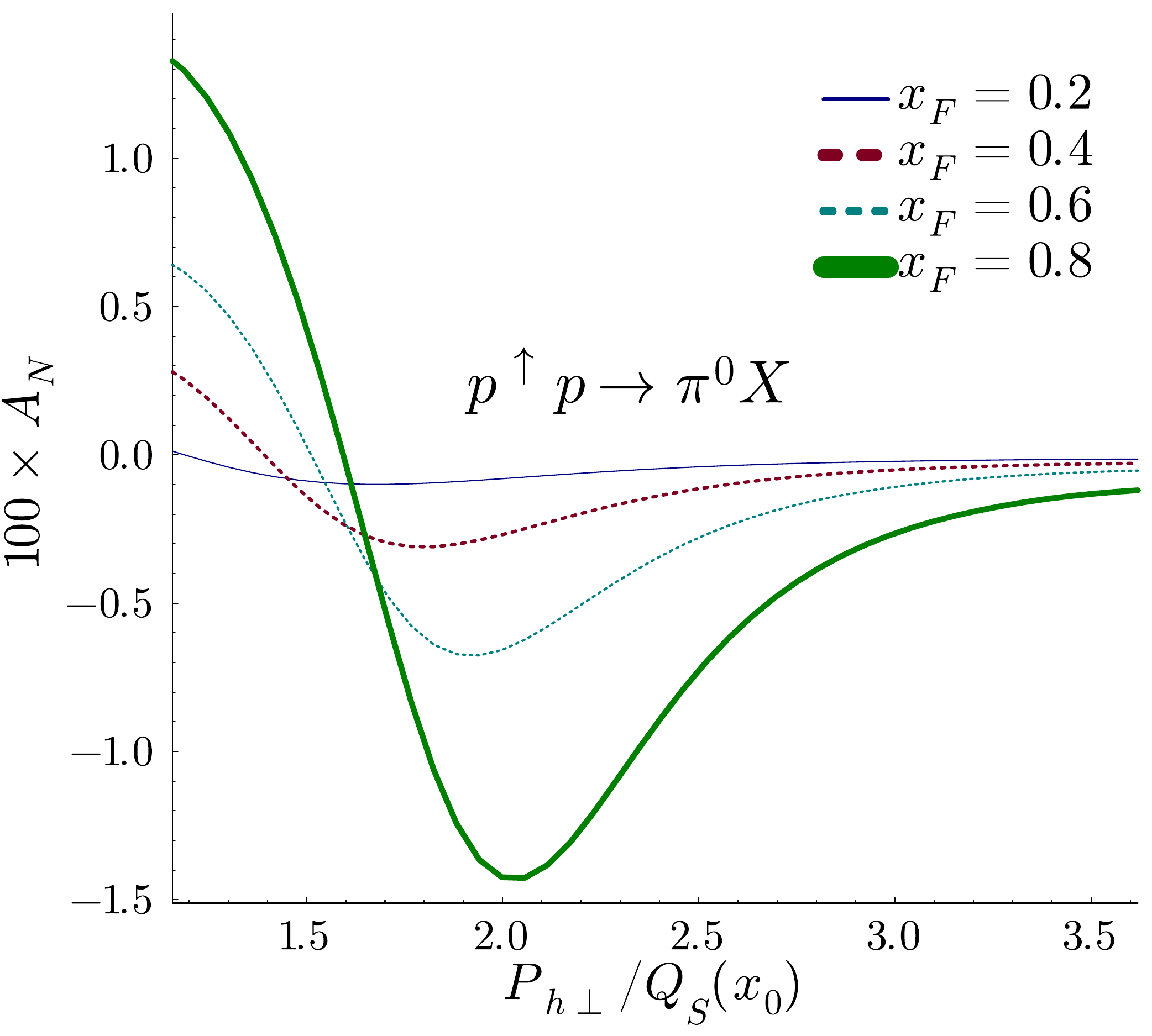}
        \includegraphics[scale=0.22]{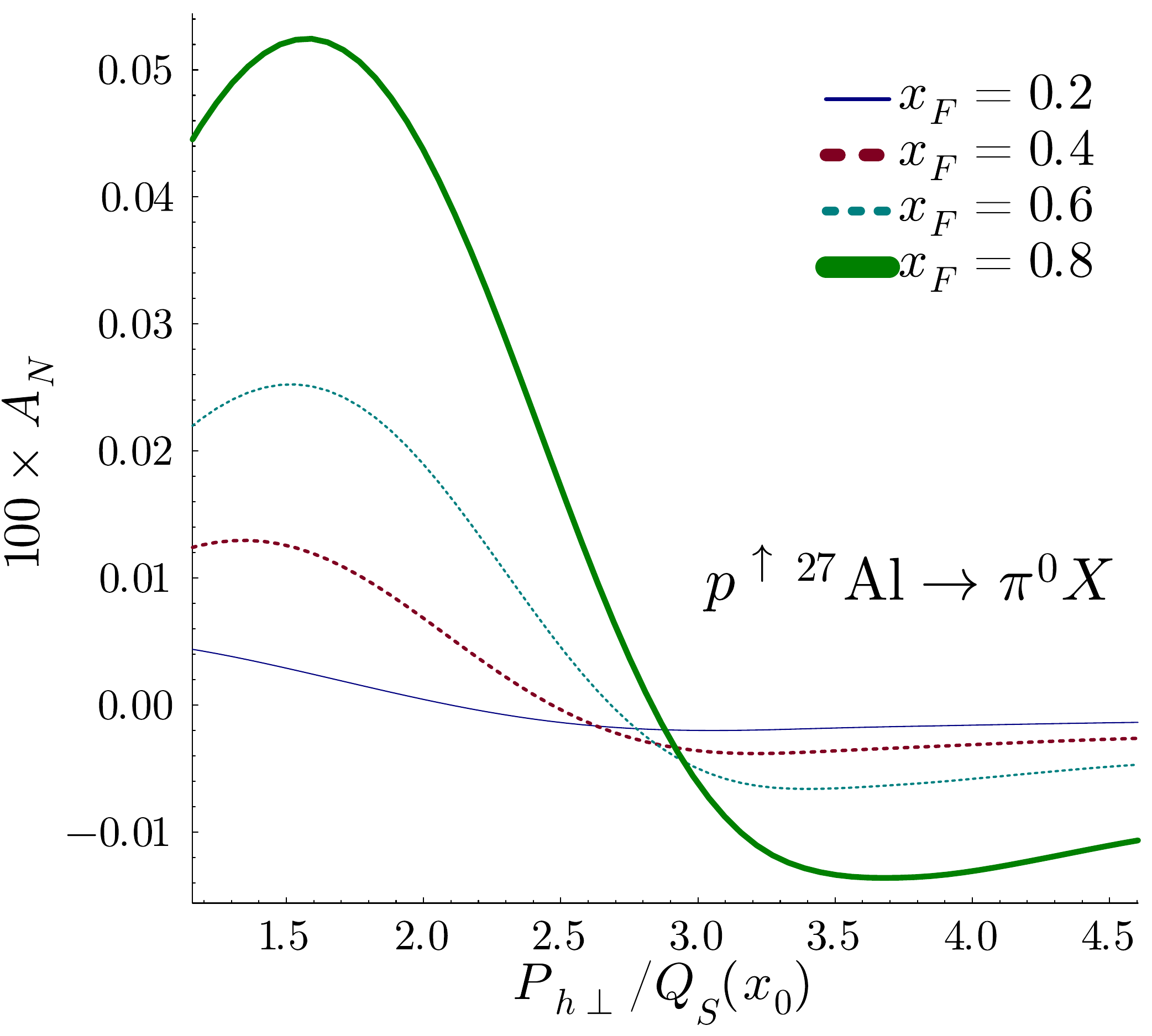}
    \caption{$A_N$ as a function of $P_{h\perp}$ for $\pi^0$ at $\sqrt{s}=200$ GeV.}
    \label{fig:phdep}
\end{figure}
On Fig.~\ref{fig:phdep} left, we show $A_N$ in $p^\uparrow p \to \pi^0 X$ production as a function of $P_{h\perp}$ for several values of $x_F$ at RHIC energy $\sqrt{s} = 200$ GeV. We find that $A_N$ changes sign as a function of $P_{h\perp}$, which is originating from the behavior of the Odderon distribution $G(x_g,k_\perp)$, see Fig.~\ref{fig:FTO}. Same result was reported in \cite{Kovchegov:2012ga} for $q A$ collisions. 
By changing $x_F$, we effectively probe the Odderon distribution at different values of $x_g$. The node position in $A_N$ is controlled by the Odderon and so almost unchanged with increasing $x_F$ - the slight shift to higher $\Php$ is due to the $z$ integration. Numerically we find $P_{h\perp} \approx 1.3 Q_S(x_0)$ (we recall: $Q_S(x_0) = 0.35$ GeV for this particular model \eqref{eqn:Pom}). Recent global analysis \cite{Casuga:2023dcf} found that the initial saturation scale is in the range $Q_S(x_0) \approx 0.39-0.62$ GeV implying a node in $A_N$ in the range $P_{h\perp} \approx 0.5-0.8$ GeV. The $A_N$ data at RHIC \cite{PHENIX:2019ouo,PHENIX:2023xnx,PHENIX:2023axd,STAR:2020grs} is taken for $P_{h\perp} \gtrsim 1.5 $ GeV, and so a lower range in $P_{h\perp}$ would be required to test this mechanism experimentally. One may benefit from changing the unpolarized target as the node location will move to a higher value: $P_{h\perp} \approx 1.3 A^{1/6} Q_S(x_0)$, albeit at the expense of reducing the magnitude of $A_N \sim A^{-7/6}$. 
On Fig.~\ref{fig:phdep}, right we show the results for $p^\uparrow \, {\rm Al} \to \pi^0 X$, where the node location is at $P_{h\perp} \approx 2.25 Q_S(x_0)$.

As $x_F$ increases, $x_g$ is decreased and so the Odderon drops in magnitude. From Fig.~\ref{fig:phdep} we find $A_N$ to be nevertheless rising with $x_F$ - the drop in the Odderon is counteracted by a stronger increase in the ratio of the transversity PDF to the unpolarized PDF. 
This model results in a percent-level asymmetry in the range $x_F \approx 0.6-0.8$ for proton targets when $P_{h\perp}$ is not too high. For nuclear targets $A_N$ is strongly suppressed. We have checked this numerically and found the numerical suppression to be very close to the parametric estimate $A_N \approx A^{-7/6}$.

The overall magnitude, as well as the behavior of $A_N$ as a function of the kinematics strongly depends on the model for the Odderon exchange and also on the twist-3 FF $\hat{e}_1(z)$. We have computed $A_N$ using two alternative models of $\hat{e}_1(z)$: the spectator model \cite{Gamberg:2003pz} and using \cite{Bauer:2022mvl} that makes the assumption that the real and the imaginary part of the dynamical twist-3 FF are same in magnitude, determining $\hat{e}_1(z)$ via \eqref{eqn:QCD_of_motion}. We find $A_N$ to be smaller in magnitude than using $\hat{e}_1(z)$ from \cite{Yuan:2003gu} and can even decrease as a function of $x_F$ in case of \cite{Gamberg:2003pz}. The node position in $A_N$ is roughly same for all models of $\hat{e}_1(z)$, and so is a robust feature of the Odderon only. This analysis confirms our numerical results rely on a specific model computation and more precise estimates depend on a future extractions of $\hat{e}_1(z)$ as well as the Odderon. In particular, the soft component of the Odderon exchange has been discovered only recently in elastic $pp$ scattering \cite{D0:2012erd,TOTEM:2018psk}. 

\section{Conclusions}
\label{sec:concl}

We found a new contribution to SSA in forward $pA$ collisions that relies on the Pomeron-Odderon interference in the dense nuclear target to generate the phase required for SSA. The result then becomes sensitive to the real part of the twist-3 FF, which is in contrast to the conventional mechanism where the imaginary part of the twist-3 FF appears. The new contribution has no counterpart in the strict collinear twist-3 framework, arising as a consequence of a higher twist description of the dense target.

Our computation predicts a node in $A_N$ as a function of $P_{h\perp}$, originating from the small-$x$ description of the Odderon in the dense target. While alternative contributions to $A_N$ can also develop a node (see, for example, \cite{Kovchegov:2020kxg}), the key point here is that the node position is fixed by the initial saturation scale and does not change strongly with $x_F$—a feature characteristic of the small-$x$ evolution of the Odderon.
For nuclear targets the node position moves to higher $P_{h\perp}$, but then $A_N$ drops strongly as $A_N \propto A^{-7/6}$. These results could be tested experimentally at RHIC or at the future LHCspin project \cite{Hadjidakis:2018ifr} with proton or light-ion targets. Pomeron-Odderon interference could also contribute to polarized $\Lambda$ production in $ep$ and $eA$ collisions 
and so this mechanism could in principle be searched for also at a future Electron Ion Collider \cite{Accardi:2012qut}.

We find it unlikely that the Pomeron-Odderon mechanism alone would be responsible for the SSA observed at RHIC. The obtained $A_N$ for forward $pp$ collision is percent level only for $x_F \gtrsim 0.6$ and in the low $P_{h\perp}$ range (around the proton saturation scale $Q_S$) - dropping to per-milles after about $P_{h\perp} \gtrsim 3-3.5 Q_S$. For $pA$ collisions, $A_N$ drops as $A_N \approx A^{-7/6}$ over a broad kinematic region and so the contribution from the real part of twist-3 FFs appears negligible for heavy targets. On the other hand, the contribution from the imaginary part of twist-3 FFs is in line \cite{Benic:2018amn} with the STAR data. Considering both of these contributions, the quantitative description of the PHENIX data still remains a challenge. Clearly more work is needed in this direction, perhaps by additionally incorporating the contribution from twist-3 distribution functions \cite{Hatta:2016wjz} and the lensing mechanism \cite{Kovchegov:2020kxg} in a comprehensive computation.

\begin{acknowledgments}
We thank Abhiram Kaushik for discussions. We acknowledge the hospitality and the support of the EIC Theory Institute at the Brookhaven National Laboratory in March 2024 and thank the Nuclear Theory group for discussions. This work is supported by the Croatian Science Foundation (HRZZ) no. 5332 (UIP-2019-04).
\end{acknowledgments}

\appendix

\section{Angular integrals}
\label{integrallll}

We first calculate angular integral from Eq.~\eqref{eqn:Hattaintegral}
\be
I=\int_0^{2\pi}\rmd\phi_k\frac{(\qop-\kp)\times\Sp}{(\qop-\kp)^2}=\int_0^{2\pi}\rmd\phi_k\frac{q_{1\perp}\sin(\phi_{Sh})-k_\perp\sin(\phi_{Sk})}{q_{1\perp}^2+k_\perp^2-2q_{1\perp} k_\perp \cos(\phi_{hk})}
\ee
where $\phi_{ab} \equiv \phi_a - \phi_b$. We have $\qonp = \Phpv/z_1$ so that $\phi_{q_1} = \phi_h$. With the help of the Appendix C from \cite{Benic:2021gya} we find
\be
\begin{split}
I &= 2\pi\sin(\phi_{Sh})\left(\frac{q_{1\perp}}{\left|q_{1\perp}^2-k_\perp^2\right|}+\frac{1}{2q_{1\perp}}\left(1-\frac{q_{1\perp}^2+k_\perp^2}{\left|q_{1\perp}^2-k_\perp^2\right|}\right)\right)=\frac{\pi\sin(\phi_{Sh})}{q_{1\perp}}\left(1+ \mathrm{sign}(q_{1\perp}-k_\perp)\right)\\
& =2\pi z_1\frac{\Phpv\times\Sp}{\Php^2}\theta\left(\frac{\Php}{z_1}-k_\perp\right)\,.
\end{split}
\ee
The $\theta$-function leads to an upper bound of the $k_\perp$ integral in \eqref{eqn:Hatta}. 

We now solve the double angular integral in the first part of \eqref{eqn:CSalm}
\be
\begin{split}
J_1 &=\int_0^{2\pi}\rmd\phi_k\int_0^{2\pi}\rmd\phi_\Delta\left[-\frac{1}{2}\frac{\Dp\times\Sp}{\left(\qonp-\kp\right)^2}+\frac{(\qonp - \kp)\times \Sp}{(\qonp - \kp)^2}\frac{(\qonp - \kp)\cdot\Dp}{(\qonp-\kp)^2}\right]\cos(\phi_{k\Delta})\,.
\end{split}
\ee
Performing the $\phi_\Delta$ integration we find
\be
\begin{split}
    J_1&=\pi\Delta_\perp\Bigg[-\frac{1}{2}\int_0^{2\pi}\rmd\phi_k\frac{\sin(\phi_{Sk})}{q_{1\perp}^2+k_\perp^2-2q_{1\perp} k_\perp \cos(\phi_{hk})}+q_{1\perp}^2\int_{0}^{2\pi}\rmd\phi_k\frac{\sin(\phi_{Sh})\cos(\phi_{hk})}{\left[q_{1\perp}^2+k_\perp^2-2q_{1\perp} k_\perp \cos(\phi_{hk})\right]^2}\\
    &-q_{1\perp}k_{\perp}\int_0^{2\pi}\rmd\phi_k\frac{\sin(\phi_{Sh})}{\left[q_{1\perp}^2+k_\perp^2-2q_{1\perp} k_\perp \cos(\phi_{hk})\right]^2} -q_{1\perp}k_\perp\int_0^{2\pi}\rmd\phi_k\frac{\sin(\phi_{Sk})\cos(\phi_{hk})}{\left[q_{1\perp}^2+k_\perp^2-2q_{1\perp} k_\perp \cos(\phi_{hk})\right]^2}\\
    & +k_\perp^2\int_0^{2\pi}\rmd\phi_k\frac{\sin(\phi_{Sk})}{\left[q_{1\perp}^2+k_\perp^2-2q_{1\perp} k_\perp \cos(\phi_{hk})\right]^2}\Bigg]\,.
\end{split}
\label{eq:J1}
\ee
Out of the five integrals in \eqref{eq:J1}, the first three and the last can be computed using the results in \cite{Benic:2021gya}. The fourth integral is found as
\be
\int_0^{2\pi}\rmd\phi_k\frac{\sin(\phi_{Sk})\cos(\phi_{h k})}{\left[q_{1\perp}^2+k_\perp^2-2q_{1\perp} k_\perp \cos(\phi_{hk})\right]^2}=\pi\sin(\phi_{Sh})\frac{|q_{1\perp}^2-k_\perp^2|^3-(q_{1\perp}^2+k_\perp^2)(q_{1\perp}^2-k_\perp^2)^2+4q_{1\perp}^2k_\perp^2(q_{1\perp}^2+k_\perp^2)}{2 q_{1\perp}^2k_\perp^2|q_{1\perp}^2-k_\perp^2|^3}
\label{eqn:remint}
\ee
We find that all the terms add up to zero
\be
\begin{split}
    J_1&=2\pi^2\Delta_\perp\sin(\phi_{Sh})\Bigg[\frac{1}{4q_{1\perp}k_\perp}\left(1-\frac{q_{1\perp}^2+k_\perp^2}{|q_{1\perp}^2-k_\perp^2|}\right)+2q_{1\perp}^3k_\perp\frac{1}{|q_{1\perp}^2-k_\perp^2|^3}-q_{1\perp}k_\perp\frac{q_{1\perp}^2+k_\perp^2}{|q_{1\perp}^2-k_\perp^2|^3}\\
    &-q_{1\perp}k_\perp\frac{|q_{1\perp}^2-k_\perp^2|^3-(q_{1\perp}^2+k_\perp^2)(q_{1\perp}^2-k_\perp^2)^2+4q_{1\perp}^2k_\perp^2(q_{1\perp}^2+k_\perp^2)}{4q_{1\perp}^2k_\perp^2|q_{1\perp}^2-k_\perp^2|^3}+2q_{1\perp}k_\perp^3\frac{1}{|q_{1\perp}^2-k_\perp^2|^3}\Bigg]=0\,.
\end{split}
\ee

The second angular integral in \eqref{eqn:CSalm} is
\be
J_2 = \int_0^{2\pi}\rmd\phi_k\int_0^{2\pi}\rmd\phi_\Delta\left[-\frac{1}{2}\frac{\Dp\times\Sp}{\left(\qonp-\kp\right)^2}+\frac{(\qonp - \kp)\times \Sp}{(\qonp - \kp)^2}\frac{(\qonp - \kp)\cdot\Dp}{(\qonp-\kp)^2}\right]\cos(\phi_{h\Delta})\,.
\ee
After performing the $\phi_\Delta$ we find:
\be
\begin{split}
    J_2&=\pi\Delta_\perp\Bigg[-\frac{1}{2}\int_0^{2\pi}\rmd\phi_k\frac{\sin(\phi_{Sh})}{q_{1\perp}^2+k_\perp^2-2q_{1\perp} k_\perp \cos(\phi_{hk})}+q_{1\perp}^2\int_0^{2\pi}\rmd\phi_k\frac{\sin(\phi_{Sh})}{\left[q_{1\perp}^2+k_\perp^2-2q_{1\perp} k_\perp \cos(\phi_{hk})\right]^2}\\
    &-q_{1\perp}k_\perp\int_0^{2\pi}\rmd\phi_k\frac{\sin(\phi_{Sh})\cos(\phi_{hk})}{\left[q_{1\perp}^2+k_\perp^2-2q_{1\perp} k_\perp \cos(\phi_{hk})\right]^2}-q_{1\perp}k_\perp\int_0^{2\pi}\rmd\phi_k\frac{\sin(\phi_{Sk})}{\left[q_{1\perp}^2+k_\perp^2-2q_{1\perp} k_\perp \cos(\phi_{hk})\right]^2}\\
    & +k_\perp^2\int_0^{2\pi}\rmd\phi_k\frac{\sin(\phi_{Sk})\cos(\phi_{ hk })}{\left[q_{1\perp}^2+k_\perp^2-2q_{1\perp} k_\perp \cos(\phi_{hk})\right]^2}\Bigg]\,.
\end{split}
\ee
Using the results from \cite{Benic:2021gya} and Eq.~\eqref{eqn:remint} we have:
\be
\begin{split}
    J_2&=\pi\Delta_\perp(2\pi)\sin(\phi_{Sh})\Bigg[-\frac{1}{2}\frac{1}{|q_{1\perp}^2-k_\perp^2|}+q_{1\perp}^2\frac{q_{1\perp}^2+k_\perp^2}{|q_{1\perp}^2-k_\perp^2|^3}-4q_{1\perp}^2k_\perp^2\frac{1}{|q_{1\perp}^2-k_\perp^2|^3}\\
    &+k_\perp^2\frac{|q_{1\perp}^2-k_\perp^2|^3-(q_{1\perp}^2+k_\perp^2)(q_{1\perp}^2-k_\perp^2)^2+4q_{1\perp}^2k_\perp^2(q_{1\perp}^2+k_\perp^2)}{4q_{1\perp}^2k_\perp^2|q_{1\perp}^2-k_\perp^2|^3}\Bigg]\\
    &= \pi\Delta_\perp(2\pi)\sin(\phi_{Sh})\frac{1+\mathrm{sign}(q_{1\perp}-k_\perp)}{4q_{1\perp}^2} =\pi^2 z_1^2 \Delta_\perp\frac{\Phpv\times\Sp}{\Php^3}\theta\left(\frac{\Php}{z_1}-k_\perp\right)\,.
\end{split}
\ee

\bibliography{references}
\end{document}